\documentclass[twocolumn,dvipsnames,twocolappendix]{aastex631}
\usepackage[T1]{fontenc}
\usepackage[utf8]{inputenc}
\usepackage[english]{babel}
\usepackage{amssymb}
\usepackage{amsmath}
\usepackage{esint}
\usepackage{dsfont}

\usepackage{bm}
\usepackage{hyperref}
\usepackage{bookmark}
\usepackage{xcolor}

\usepackage{tikz}
\usetikzlibrary{calc, shapes, shapes.geometric, arrows, fit, backgrounds, positioning, decorations.pathreplacing, decorations.pathmorphing, decorations.shapes, calligraphy}

\usepackage{xcolor}

\definecolor{red}{HTML}{BF616A}
\definecolor{orange}{HTML}{D08770}
\definecolor{yellow}{HTML}{EBCB8B}
\definecolor{green}{HTML}{A3BE8C}
\definecolor{purple}{HTML}{B48EAD}

\DeclareMathAlphabet{\mathup}{OT1}{\familydefault}{m}{n}
\newcommand\diff{\mathop{}\!\mathup{d}}

\usepackage[shortcuts]{extdash}

\newcommand\hd[1]{HD~#1}
\newcommand\hip[1]{HIP~#1}

\newcommand\flatval{0.889}
\newcommand\protval{15.3}
\newcommand\protvaldr{18.5}

\begin{document}
\title{A long spin period for a sub-Neptune-mass exoplanet}

\author[0000-0002-3286-3543]{Ellen M. Price}
\altaffiliation{Heising-Simons Foundation 51 Pegasi b Postdoctoral Fellow}
\affiliation{Department of the Geophysical Sciences, University of Chicago, 5734 S Ellis Avenue, Chicago, IL 60637, USA}

\author[0000-0002-7733-4522]{Juliette Becker}
\affiliation{Department of Astronomy, University of Wisconsin-Madison, 475 N Charter Street, Madison, WI 53706, USA}

\author[0000-0002-7564-6047]{Zo\"e L. de Beurs}
\altaffiliation{NSF Graduate Research Fellow, MIT Presidential Fellow, MIT Collamore-Rogers Fellow, MIT Teaching Development Fellow}
\affiliation{Department of Earth, Atmospheric and Planetary Sciences, Massachusetts Institute of Technology, 77 Massachusetts Avenue, Cambridge, MA 02139, USA}

\author[0000-0003-0638-3455]{Leslie A. Rogers}
\affiliation{Department of Astronomy \& Astrophysics, University of Chicago, 5640 S Ellis Avenue, Chicago, IL 60637, USA}

\author[0000-0001-7246-5438]{Andrew Vanderburg}
\altaffiliation{Alfred P. Sloan Research Fellow}
\affiliation{Department of Physics and Kavli Institute for Astrophysics and Space Research, Massachusetts Institute of Technology, 77 Massachusetts Avenue, Cambridge, MA 02139, USA}

\begin{abstract}
\hip{41378~f} is a sub-Neptune exoplanet with an anomalously low density. Its long orbital period and deep transit make it an ideal candidate for detecting oblateness photometrically. We present a new cross-platform, GPU-enabled code \texttt{greenlantern}, suitable for computing transit light curves of oblate planets at arbitrary orientations. We then use Markov Chain Monte Carlo to fit K2 data of \hip{41378~f}, specifically examining its transit for evidence of oblateness and obliquity. We find that the flattening of \hip{41378~f} is $f \leq \flatval$ at the 95\% confidence level, consistent with a rotation period of $P_\text{rot} \geq \protval$~hr. In the future, high-precision data from JWST has the potential to tighten such a constraint and can differentiate between spherical and flattened planets.
\end{abstract}


\section{Introduction}

\noindent The distribution of angular momentum within a planetary system, including both orbital and rotational angular momentum for each body, serves as a valuable dynamical fossil, helping us to understand the formation and evolution of the system \citep[i.e.,][]{Ward2004, Hamilton2004, Saillenfest2021a}. One of the most challenging measurements to obtain is the rotational angular momentum of a planet. The magnitude of the angular momentum is set by physical planet parameters (the planetary mass, radius, and interior density profile), which are largely fixed at formation, and the planetary rotation rate, which may evolve due to dynamical interactions post-formation. The direction of the angular momentum, referred to as the planetary obliquity, also provides insight into the planet's formation and subsequent dynamical evolution \citep[e.g.,][]{Slattery1992, Laskar1993, Millholland2024} and can, in some cases, improve estimates of a planet's potential for habitability \citep{Quarles2020, Vervoort2022}.

The non-sphericity of a planetary body can also be used to infer its rotational angular momentum. Planets with short rotational periods may become oblate due to rapid rotation, with faster rotators experiencing greater deformation \citep{Seager2002}. At the same time, the planet's obliquity determines how the planet's rotational axis is oriented relative to the observer's line of sight. 
For directly imaged planets, a planet's obliquity and rotation can be measured through direct spectroscopy, an estimate of the photometric rotation period, and an estimate of the planet's radius \citep[e.g.,][]{Bryan2020, Bryan2021}. 
For planets seen in transit, the resulting light curve of a oblate body will reflect a flattened shape, rather than the commonly-assumed spherical transiting body. This shape is often parameterized in terms of a flattening $f$, as in 
\citet{BarnesFortney2003}: 
\begin{equation}
    f \equiv \frac{R_\text{eq} - R_\text{pol}}{R_\text{eq}},
\end{equation}
where $R_\text{eq}$ is the planet's equatorial radius and $R_\text{pol}$ is its polar radius. 
\citet{BarnesFortney2003} present a case study on \hd{209458~b}, demonstrating that planetary oblateness introduces a deviation from the spherical light curve. While this deviation is small --- about 10~ppm --- it can bias the accurate recovery of other transit parameters if flattening is not accounted for in the fits \citep{Berardo+2022}. 
Precise instrumentation, long observation baselines, and appropriate oblate transit models allow the possibility of detecting such a signal and subsequently constraining planetary oblateness and obliquity.
However, due to the difficulty of the measurement of such a small signal, only a few detections and constraints of such deformation have been made to date, and only for Jupiter-sized planets where the signal is more pronounced \citep{Zhu2014, Biersteker2017, Barros2022, Akinsanmi2024}. 


The transit fits used by many planet discovery and characterization papers assume spherical bodies --- for example, publications using \texttt{exofast} \citep{Eastman+2013} and \texttt{batman} \citep{Kreidberg2015} --- since computing a light curve using a non-spherical model is significantly more computationally expensive. 
Despite this computational limitation, there have been previous efforts to build codes to model the transit of an ellipsoidal body. 
Initial work by \citet{BarnesFortney2003} use a numerical integration scheme based on Romberg's method to compute the fraction of light blocked by an orbiting ellipse, weighted by the stellar intensity. This approach effectively requires two levels of integration, which is less efficient than the approach taken by \citet{Pal2012} for spherical planets, which leverages Green's theorem to reduce the dimensionality of the integral. Other works, such as \citet{CarterWinn2010a}, \citet{Carado+2020}, and \citet{Berardo+2022} use Monte Carlo integration to evaluate the area of the projected ellipsoid that lies within the stellar disk. 
The \texttt{ellc} code \citep{Maxted2016} can compute the flux from an ellipsoid planet orbiting an ellipsoid star, though it does require accurately identifying intersection points of two ellipses. The \texttt{starry} code \citep{Luger+2021}, in its latest release, can also handle oblate maps.
\citet{Leconte+2011} consider phase curves of ellipsoid bodies, but that approach cannot be directly applied to transit light curves: For the phase curve, only the projected area is needed, but transit modeling requires the functional form of the boundary to handle its intersection with the stellar disk during transit.

In this work, we present a publicly-available\footnote{The full source code is available at \url{https://github.com/emprice/greenlantern} and identified by the DOI \texttt{10.5281/zenodo.14510631}}., GPU-accelerated\footnote{This code will run on any device that supports the OpenCL standard, which includes many CPUs. GPU acceleration is still recommended for best performance.}, and cross-platform code \texttt{greenlantern}, which combines the advantages of the \citet{Pal2012} methodology with ellipsoid planet geometry to compute transit light curves of ellipsoids at any orientation.
This code will allow for more efficient calculation of ellipsoidal transit models. We demonstrate its functionality by applying our new code to the light curve of \hip{41378~f}, a cold sub-Neptune, to constrain its rotation rate. \hip{41378~f} in an excellent target for such a constraint due to its large radius, bright host star, and low bulk density. 
At the same time, \hip{41378~f} is a planet where a rotation rate would provide significant dynamical insight: Formation models remain uncertain on how to explain the low bulk density \citep{Santerne2019} and low inferred core mass \citep{Belkovski2022} of such a cold, Jupiter-radius planet. These investigations would be aided by a constraint on \hip{41378~f}'s rotation.

This paper is structured as follows. In Section~\ref{sec:ellipsoid-transit}, we present our analytic model for the projected profile of a triaxial ellipsoid and the computational method that computes the associated transit light curve. In Section~\ref{sec:data-and-fitting}, we present the dataset for \hip{41378~f} and our method for fitting the model to this data. Section~\ref{sec:results} describes the constraints we derive on the flattening and obliquity of \hip{41378~f}, and the implications of our findings are discussed in Section~\ref{sec:discussion}. Finally, we conclude in Section~\ref{sec:conclusions}.

\section{Ellipsoid Transit Model}
\label{sec:ellipsoid-transit}

\noindent We work in a Cartesian coordinate system where the $\hat{\bm{z}}$ axis points along the line of sight from the observer to the host star. The $\hat{\bm{y}}$ axis is oriented along the ``horizontal'' transit chord, and the $\hat{\bm{x}}$ axis is oriented in the remaining perpendicular direction, so that the resulting coordinate system is right-handed. To define the orientation of an orbiting triaxial ellipsoid, we require five rotation angles. Two angles set the orbital position of the planet: $\alpha$ plays the role of mean anomaly (assuming zero eccentricity) and $\beta$ is an elevation angle related to the typical inclination $i$ by $\beta = 90^\circ - i$. We additionally define a set of primed coordinates, $\hat{\bm{x}}^\prime$, $\hat{\bm{y}}^\prime$, and $\hat{\bm{z}}^\prime$, along the three semiaxes of the ellipsoid, which have lengths $a$, $b$, and $c$, respectively. Three angles $\zeta$, $\eta$, and $\xi$ set the orientation of the ellipsoid in space: $\zeta$ rotates around $\hat{\bm{x}}^\prime$, $\eta$ rotates around $\hat{\bm{y}}^\prime$, and $\xi$ rotates around $\hat{\bm{z}}^\prime$. To move between these two coordinate systems, we use the mapping
\begin{equation}
    \begin{pmatrix} x^\prime \\ y^\prime \\ z^\prime \end{pmatrix} = R\!\left(\zeta, \eta, \xi\right) \left[\begin{pmatrix} x \\ y \\ z \end{pmatrix} + R_y\!\left(-\beta\right) R_x\!\left(\alpha\right) \begin{pmatrix} 0 \\ 0 \\ d \end{pmatrix}\right]
    \label{eqn:coord-map}
\end{equation}
where $d$ is the distance from the ellipsoid's origin to the center of the star, and
\begin{equation}
    R_x\!\left(\theta\right) = \begin{pmatrix} 
        1 & 0 & 0 \\
        0 & \cos{\theta} & -\sin{\theta} \\
        0 & \sin{\theta} & \cos{\theta}
    \end{pmatrix},
\end{equation}
\begin{equation}
    R_y\!\left(\theta\right) = \begin{pmatrix}
        \cos{\theta} & 0 & \sin{\theta} \\
        0 & 1 & 0 \\
        -\sin{\theta} & 0 & \cos{\theta}
    \end{pmatrix},
\end{equation}
and
\begin{equation}
    R_z\!\left(\theta\right) = \begin{pmatrix}
        \cos{\theta} & -\sin{\theta} & 0 \\
        \sin{\theta} & \cos{\theta} & 0 \\
        0 & 0 & 1
    \end{pmatrix}
\end{equation}
are the typical rotation matrices about $x$, $y$, and $z$, respectively. To simplify notation, we define the product
\begin{equation}
    R\!\left(\zeta, \eta, \xi\right) \equiv R_x\!\left(\zeta\right) R_y\!\left(\eta\right) R_z\!\left(\xi\right) = \begin{pmatrix}
        r_1 & r_2 & r_3 \\
        r_4 & r_5 & r_6 \\
        r_7 & r_8 & r_9
    \end{pmatrix}
\end{equation}
whose components appear in equations below. In Figure~\ref{fig:geometry}, we illustrate the geometry of the global coordinate system and its associated angles.

\begin{figure*}
    \centering
    \resizebox{\linewidth}{!}{%
        \begin{tikzpicture}[scale=2]%
            \begin{scope}[shift={(0,0)}]%
                \begin{scope}[shift={(-1.5,-0.5)}]%
                    \draw[very thick, black, ->] (0,0) -- node[anchor=west, pos=1] {\Large $\hat{\bm{y}}$} +(1,0);%
                    \draw[very thick, black, ->] (0,0) -- node[anchor=south, pos=1] {\Large $\hat{\bm{z}}$} +(0,1);%
                    \draw[thick, black] (0,0) circle (0.075);%
                    \node[thick, black, anchor=north east] at (0,0) {\Large $\hat{\bm{x}}$};%
                \end{scope}%
                \begin{scope}[shift={(0,3)}]%
                    \begin{scope}[rotate around={30:(0,0)}, shift={(0,-3)}]%
                        \draw[line width=2pt, purple] (0,0) ellipse[x radius=1, y radius=0.8, rotate=-20];%
                        \begin{scope}[shift={(0.5,0)}, rotate around={-20:(0,0)}]%
                            \draw[very thick, purple!50!black, ->] (0,0) -- node[anchor=west, pos=1] {\Large $\hat{\bm{y}}^\prime$} +(1,0);%
                            \draw[very thick, purple!50!black, ->] (0,0) -- node[anchor=south, pos=1] {\Large $\hat{\bm{z}}^\prime$} +(0,1);%
                            \draw[very thick, purple!50!black, ->] (0,0) -- node[anchor=north, pos=1] {\Large $\hat{\bm{x}}^\prime$} +(-0.25,-0.25);%
                        \end{scope}%
                    \end{scope}%
                    \draw[line width=3pt, yellow] (0,0) ++(-2,0) -- ++(4,0);%
                    \draw[thick, black, dashed] (0,0) -- ++(0,-2);%
                    \draw[thick, black, -*] (0,0) -- node[anchor=south west, pos=0.5] {\Large $d$} ++(-60:3);%
                    \draw[thick, black, ->] (0,0) ++(0,-1) arc (90:120:-1) node[anchor=north, pos=0.4] {\Large $\alpha$};%
                \end{scope}%
            \end{scope}%
            \begin{scope}[shift={(4.5,0)}]%
                \begin{scope}[shift={(-0.5,-0.5)}]%
                    \draw[very thick, black, ->] (0,0) -- node[anchor=west, pos=1] {\Large $\hat{\bm{z}}$} +(1,0);%
                    \draw[very thick, black, ->] (0,0) -- node[anchor=south, pos=1] {\Large $\hat{\bm{x}}$} +(0,1);%
                    \draw[thick, black] (0,0) circle (0.075);%
                    \node[thick, black, anchor=north east] at (0,0) {\Large $\hat{\bm{y}}$};%
                \end{scope}%
                \begin{scope}[shift={(0,3)}]%
                    \begin{scope}[rotate around={-30:(0,0)}]%
                        \draw[line width=2pt, purple] (0,0) ellipse[x radius=0.8, y radius=0.85, rotate=10];%
                        \begin{scope}[shift={(0,0.5)}, rotate around={10:(0,0)}]%
                            \draw[very thick, purple!50!black, ->] (0,0) -- node[anchor=west, pos=1] {\Large $\hat{\bm{z}}^\prime$} +(1,0);%
                            \draw[very thick, purple!50!black, ->] (0,0) -- node[anchor=south, pos=1] {\Large $\hat{\bm{x}}^\prime$} +(0,1);%
                            \draw[very thick, purple!50!black, ->] (0,0) -- node[anchor=north east, pos=1] {\Large $\hat{\bm{y}}^\prime$} +(-0.25,-0.25);%
                        \end{scope}%
                    \end{scope}%
                    \draw[line width=3pt, yellow] (0,0) ++(-30:3) ++(0,-2) -- ++(0,4);%
                    \draw[thick, dashed, black] (0,0) ++(-30:3) -- ++(-2,0);%
                    \draw[thick, black, ->] (0,0) ++(-30:3) +(-1,0) arc (0:-30:-1) node[anchor=south east, pos=0.2] {\Large $\beta$};%
                    \draw[thick, black, *-] (0,0) -- node[anchor=south west, pos=0.5] {\Large $d$} ++(-30:3);%
                \end{scope}%
            \end{scope}%
            \begin{scope}[shift={(8,0)}]%
                \begin{scope}[shift={(0,-0.5)}]%
                    \draw[very thick, black, ->] (0,0) -- node[anchor=west, pos=1] {\Large $\hat{\bm{y}}$} +(1,0);%
                    \draw[very thick, black, ->] (0,0) -- node[anchor=south, pos=1] {\Large $\hat{\bm{x}}$} +(0,1);%
                    \draw[thick, black] (0,0) ++(0.05,0.05) -- ++(-0.1,-0.1);%
                    \draw[thick, black] (0,0) ++(0.05,-0.05) -- ++(-0.1,0.1);%
                    \node[thick, black, anchor=north east] at (0,0) {\Large $\hat{\bm{z}}$};%
                \end{scope}%
                \begin{scope}[shift={(2.5,1.5)}]%
                    \draw[line width=3pt, yellow] (0,0) circle (2);%
                    \begin{scope}[shift={(2,1)}]%
                        \begin{scope}[rotate around={20:(0,0)}]%
                            \draw[line width=2pt, purple] (0,0) ellipse[x radius=1, y radius=0.9, rotate=15];%
                            \begin{scope}[shift={(0.5,0.5)}, rotate around={15:(0,0)}]%
                                \draw[very thick, purple!50!black, ->] (0,0) -- node[anchor=west, pos=1] {\Large $\hat{\bm{y}}^\prime$} +(1,0);%
                                \draw[very thick, purple!50!black, ->] (0,0) -- node[anchor=south, pos=1] {\Large $\hat{\bm{x}}^\prime$} +(0,1);%
                                \draw[very thick, purple!50!black, ->] (0,0) -- node[anchor=north east, pos=1] {\Large $\hat{\bm{z}}^\prime$} +(-0.25,-0.25);%
                            \end{scope}%
                        \end{scope}%
                    \end{scope}%
                \end{scope}%
            \end{scope}%
        \end{tikzpicture}%
    }%
    \caption{Illustration of the coordinate system and rotation angles used to model the transit of an ellipsoidal planet. The observer's line of sight aligns with the $\hat{\bm{z}}$ axis and the $\hat{\bm{y}}$ axis is oriented along the horizontal transit chord; the $\hat{\bm{x}}$ axis is mutually perpendicular to $\hat{\bm{y}}$ and $\hat{\bm{z}}$ such that the coordinate system follows the right-hand rule. The ellipsoidal planet's orbit is described by two rotation angles: $\alpha$ (rotation around $\hat{\bm{x}}$) and $\beta$ (rotation around $\hat{\bm{y}}$), which correspond to the mean anomaly and elevation angle, respectively. The orbital distance $d$ is measured to the stellar center; though the star is spherical by assumption, we collapse its geometry to a plane for simplicity. The left panel shows the rotation of the ellipsoid by $\alpha$, the middle panel illustrates rotation by $\beta$, and the right panel depicts the view by an observer. The additional angles $\zeta$, $\eta$, and $\xi$ are not shown, but their associated axes $\hat{\bm{x}}^\prime$, $\hat{\bm{y}}^\prime$, and $\hat{\bm{z}}^\prime$ are drawn as a projection.}
    \label{fig:geometry}
\end{figure*}

Computing the projection of an ellipsoid onto the $x$--$y$ plane can be done with a computer algebra system, though simplifying the resulting expressions into a compact and usable form is tedious. Here, we present one such simplification. Solving for the bounding curve of the ellipsoid projection, we find the parametric representation
\begin{align}
    x_\text{e} &= x_0 + \delta x \cos{\theta} \\
    y_\text{e} &= y_0 + \delta y_1 \cos{\theta} + \delta y_2 \sin{\theta}
\end{align}
in terms of an angle $\theta$, where
\begin{equation}
    x_0 = d \cos{\alpha} \sin{\beta},
\end{equation}
\begin{equation}
    y_0 = d \sin{\alpha},
\end{equation}
\begin{equation}
    \delta x = \sqrt{a^2 r_1^2 + b^2 r_4^2 + c^2 r_7^2},
    \label{eqn:dx}
\end{equation}
\begin{equation}
    \delta y_1 = \left(a^2 r_1 r_2 + b^2 r_4 r_5 + c^2 r_7 r_8\right) \Big/ \delta x,
    \label{eqn:dy1}
\end{equation}
and
\begin{equation}
    \delta y_2 = \sqrt{a^2 b^2 r_9^2 + a^2 c^2 r_6^2 + b^2 c^2 r_3^2} \Big/ \delta x.
    \label{eqn:dy2}
\end{equation}

We follow the approach of \citet{Pal2012} in leveraging Green's theorem to convert an integral over area within this closed curve into a line integral, which we parameterize by $\theta$; this approach also allows us to account for the effects of stellar limb darkening. We require the quantities
\begin{align}
    \frac{\partial x_\text{e}}{\partial \theta} &= -\delta x \sin{\theta} \\
    \frac{\partial y_\text{e}}{\partial \theta} &= -\delta y_1 \sin{\theta} + \delta y_2 \cos{\theta}
\end{align}
to compute the line integrals.

We have not yet addressed how to handle the case when the bounding curve intersects --- or falls completely outside --- the stellar disk, however, and this becomes important for evaluating these line integrals, since the desired area is the intersection of the stellar disk with the boundary of the projected ellipsoid. Solving for the points of intersection of the projected ellipsoid curve with the stellar disk, assumed to be a circle of unit radius, is unnecessary and actually detrimental to the accurate and efficient evaluation of the line integral. Instead, if a point on the projected ellipsoid curve $\left(x_\text{e}, y_\text{e}\right)$ falls outside the stellar disk, $x_\text{e}^2 + y_\text{e}^2 > 1$, we ``snap'' the coordinate to the coordinate along the same ray from the center of the stellar disk that falls on the edge of the stellar disk, $\left(x_\text{d}, y_\text{d}\right)$; this is illustrated in Figure~\ref{fig:snap}. We compute an angle $\lambda$ such that
\begin{equation}
    \lambda \equiv \tan^{-1}{y_\text{e} / x_\text{e}}
\end{equation}
and then have
\begin{align}
    x_\text{d} &= \cos{\lambda}, &
    y_\text{d} &= \sin{\lambda},
\end{align}
\begin{align}
    \frac{\partial x_\text{d}}{\partial \theta} &= -y_\text{d} \frac{\partial \lambda}{\partial \theta}, &
    \frac{\partial y_\text{d}}{\partial \theta} &= x_\text{d} \frac{\partial \lambda}{\partial \theta},
\end{align}
where
\begin{equation}
    \frac{\partial \lambda}{\partial \theta} = \left(-y_\text{e} \frac{\partial x_\text{e}}{\partial \theta} + x_\text{e} \frac{\partial y_\text{e}}{\partial \theta}\right) \Big/ \left(x_\text{e}^2 + y_\text{e}^2\right).
\end{equation}
When this point $\left(x_\text{d}, y_\text{d}\right)$ is used instead of $\left(x_\text{e}, y_\text{e}\right)$, we can rewrite the line integral, now parameterized by $\lambda$, to parameterize by $\theta$ again, since
\begin{equation}
    \oint\limits_\text{d} f\!\left(x, y\right) \diff\lambda = \oint\limits_\text{d} f\!\left(x, y\right) \frac{\partial \lambda}{\partial \theta} \diff \theta.
\end{equation}

\begin{figure}
    \centering
    \resizebox{\linewidth}{!}{%
        \begin{tikzpicture}[scale=2]%
            \node (scenter) at (0,0) {};%
            \draw[line width=3pt, yellow] (scenter) ++(-10:2) arc (-10:70:2);%
            \node (pcenter) at (2,1) {};%
            \draw[line width=2pt, purple, rotate around={20:(pcenter)}] (pcenter) ellipse (1 and 0.8);%
            \node[anchor=center, inner sep=0, outer sep=0] (theta0) at ($(pcenter) +(10:1 and 0.8)$) {};%
            \node[anchor=center, inner sep=0, outer sep=0, rotate around={20:(pcenter)}] (theta1) at ($(pcenter) +(70:1 and 0.8)$) {};%
            \node[anchor=south] at (theta1) {$\left(x_\text{e}, y_\text{e}\right)$};%
            \draw[thick, black, dashed] (pcenter) -- (theta0);%
            \draw[thick, black, shorten >= -3pt, shorten <= -3pt, -*] (pcenter) -- (theta1) node[anchor=center, inner sep=0, outer sep=0] (theta2) {};%
            \node (theta3) at ($(pcenter)!0.5!(theta0)$) {};%
            \node (theta4) at ($(pcenter)!0.5!(theta1)$) {};%
            \draw[thick, black, ->, shorten <= -3pt, shorten >= -3pt] (theta3) to[bend right] node[anchor=south, pos=0.5] {$\theta$} (theta4);%
            \draw[line width=2pt, dashed, purple] (scenter) ++(5:2) arc (5:50:2);%
            \draw[thick, black, dashed] (scenter) -- node (xi0) {} +(2,0);%
            \node (xi1) at ($(scenter)!0.75!(theta2)$) {};%
            \node[anchor=east, outer sep=4pt] at (xi1) {$\left(x_\text{d}, y_\text{d}\right)$};%
            \draw[thick, black, shorten <= -2pt, shorten >= -4pt, -*] (scenter) -- (xi1);%
            \draw[thick, black, dashed, shorten <= -6pt] (xi1) -- (theta2);%
            \node (xi2) at ($(scenter)!0.6!(xi0)$) {};%
            \node (xi3) at ($(scenter)!0.3!(xi1)$) {};%
            \draw[thick, black, ->, shorten <= -4pt, shorten >= -4pt] (xi2) to[bend right] node[anchor=west, pos=0.5] {$\lambda$} (xi3);%
        \end{tikzpicture}%
    }%
    \caption{Diagram of the coordinate ``snapping'' described in the text. The point $\left(x_\text{e}, y_\text{e}\right)$ on the boundary of the projected ellipsoid falls outside the stellar disk, so the line integral should be carried out on the dashed purple curve instead of the solid purple curve. This point can be mapped to the point on the disk edge at the same angular displacement $\lambda$, $\left(x_\text{d}, y_\text{d}\right)$.}
    \label{fig:snap}
\end{figure}

For a quadratic stellar limb darkening model, we need to evaluate the integrals of three functions, derived by \citet{Pal2012},
\begin{align}
    f_\text{flat} &= \frac{1}{2} \left(x \frac{\partial y}{\partial \theta} - y \frac{\partial x}{\partial \theta}\right), \\
    f_\text{lin} &= \frac{2}{3} \left[1 - \left(1 - x^2 - y^2\right)^{3/2}\right] f_\text{flat} \Big/ \left(x^2 + y^2\right), \\
    f_\text{quad} &= \frac{1}{2} \left[x \left(\frac{x^2}{3} + y^2\right) \frac{\partial y}{\partial \theta} - y \left(x^2 + \frac{y^2}{3}\right) \frac{\partial x}{\partial \theta}\right],
\end{align}
over the closed curve encompassing the area of the projected ellipsoid that is also within the stellar disk; here, $f_\text{flat}$ is a constant contribution to the intensity, $f_\text{lin}$ is a linear contribution, and $f_\text{quad}$ is a quadratic contribution. In these integrands, $\left(x, y\right)$ are either $\left(x_\text{e}, y_\text{e}\right)$ when $x_\text{e}^2 + y_\text{e}^2 < 1$ or $\left(x_\text{d}, y_\text{d}\right)$ when $x_\text{e}^2 + y_\text{e}^2 > 1$. Computationally, the latter case depends on quantities computed for the former, so only one branching instruction is needed to modify the values used in evaluating this integrand. The final value of each integral is computed numerically using Simpson's rule.

Minimizing branching this way allows us to leverage the massive parallelization available on graphics processing units (GPUs), and each sum can be computed efficiently using hierarchical parallelism on GPU. Solving for the intersections of the ellipsoid projection with the stellar disk, as done by \citet{Liu+2024}, introduces so much additional branching that we observed degraded performance with that approach on GPU, in addition to numerical instability if roots were not found to high enough precision. We use single precision floating point, along with a few measures for mitigating numerical noise, since these can have better than twice the efficiency of double precision floating point on GPUs. For example, the NVIDIA Quadro RTX 5000 card used in our calculations has a peak 32-to-1 performance ratio favoring single precision over double precision.

\subsection{Model Validation}

\begin{figure}
    \centering
    \includegraphics[width=\linewidth]{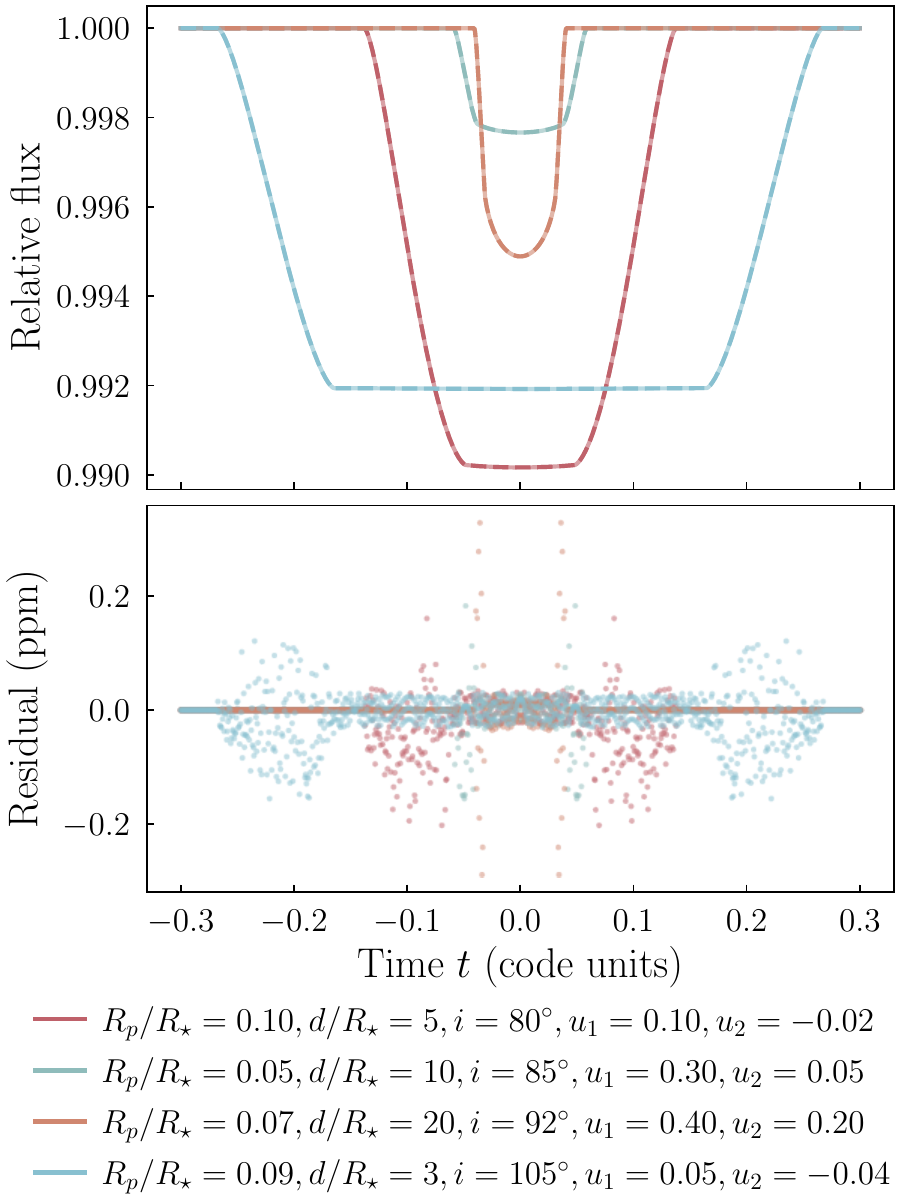}
    \caption{Gallery of several validation tests performed against \texttt{batman} \citep{Kreidberg2015} showing the computed light curve (top panel) and absolute residual (bottom panel); the parameters for each test are given in the key below the plot. In the top panel, the light, solid curves are output from \texttt{batman}, and the dark, dashed curves are output from \texttt{greenlantern}, showing qualitative and quantitative agreement. The maximum absolute residual is encountered during ingress and egress and does not exceed 0.5~ppm.}
    \label{fig:validation-with-batman}
\end{figure}

\noindent A first test to demonstrate the accuracy of our method and implementation is comparing the flux predicted by our model for a spherical shape, where the semi-axis lengths of the ellipsoid $a = b = c$, to that of an established code, \texttt{batman} \citep{Kreidberg2015}. In Figure~\ref{fig:validation-with-batman}, we provide several example light curves computed with both codes and the residuals between them, showing good qualitative and quantitative agreement at sub-ppm levels. Some noise is expected since our code computes in single precision floating point, but the amplitude does not exceed 0.5~ppm in any of the demonstrations.

\begin{figure}
    \centering
    \includegraphics[width=\linewidth]{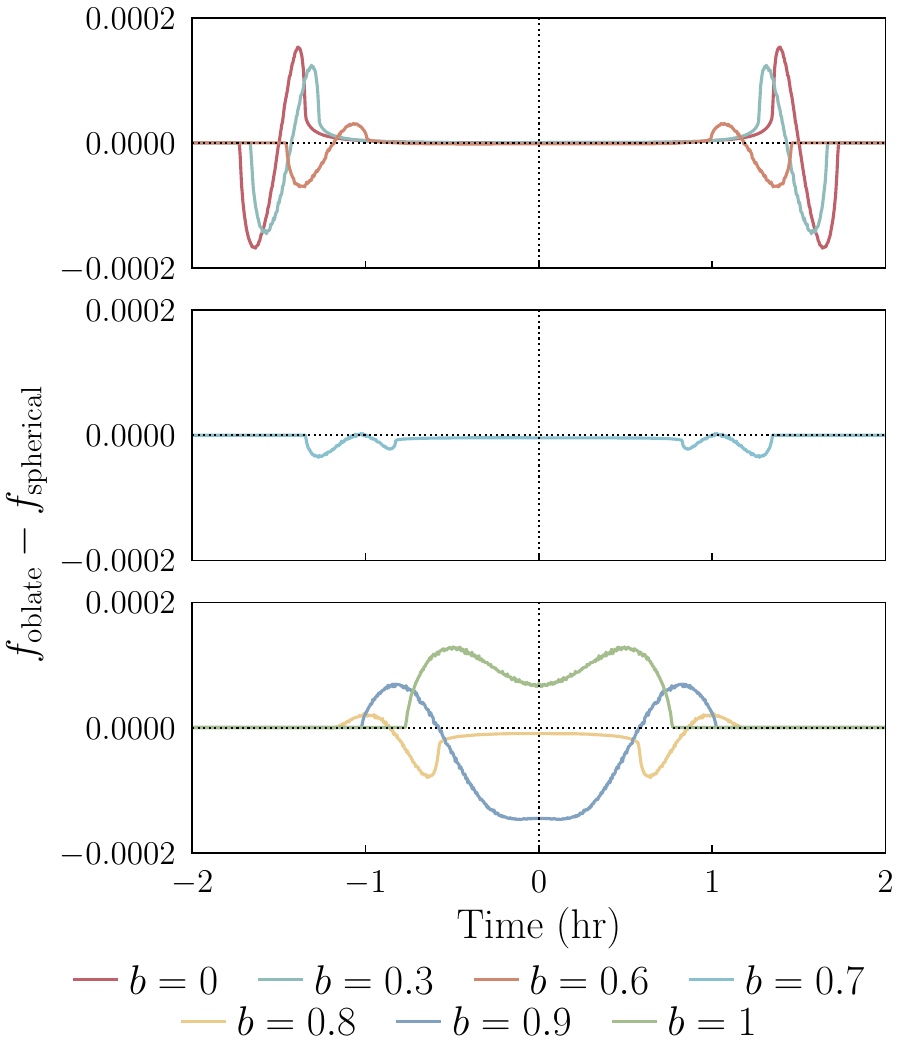}
    \caption{Analog to Figure~4 of \citet{BarnesFortney2003}, showing the difference between the relative flux of an oblate transit model and the spherical counterpart of the same cross-sectional area at mid-transit and $b = 0$, for various values of impact parameter $b$. The small-amplitude numerical noise visible in the plot is expected since the computations are carried out in single precision floating point.}
    \label{fig:barnes-fortney-fig4}
\end{figure}

\begin{figure}
    \centering
    \includegraphics[width=\linewidth]{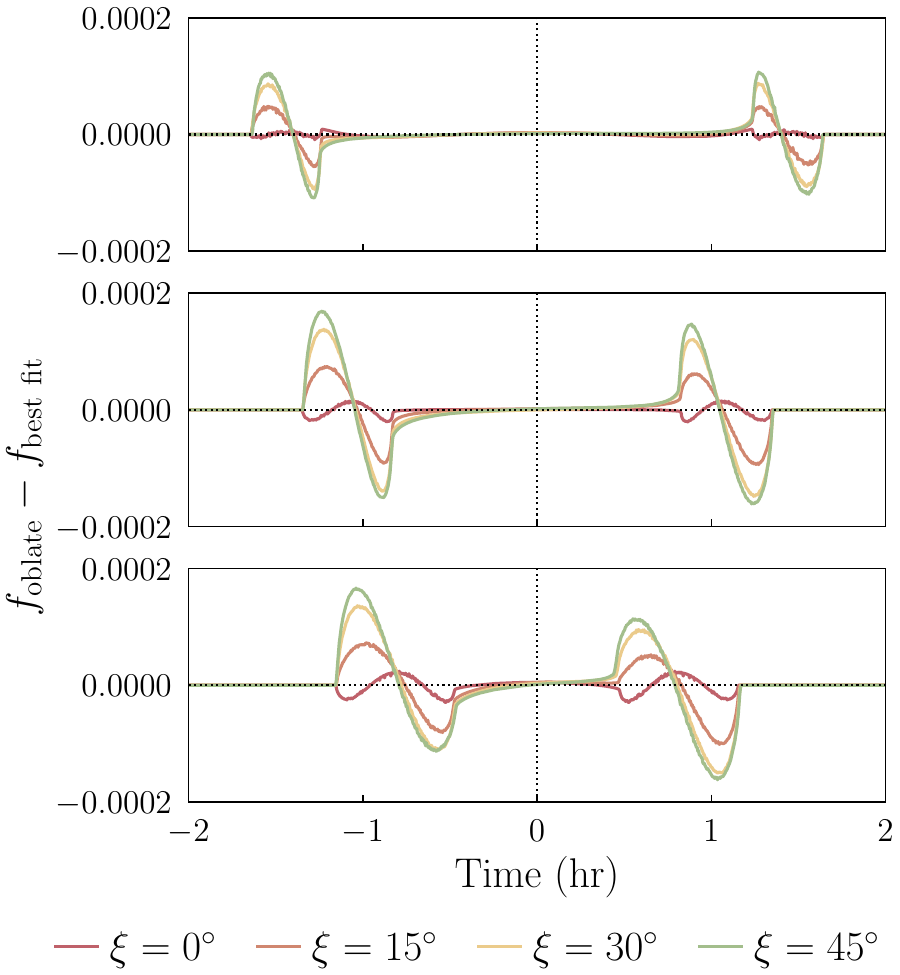}
    \caption{Analog to Figure~9 of \citet{BarnesFortney2003}, showing the difference between the relative flux of an oblate transit model and the best fit spherical model, for various values of orientation angle $\xi$ and impact parameter $b$. The small-amplitude numerical noise visible in the plot is expected since the computations are carried out in single precision floating point.}
    \label{fig:barnes-fortney-fig9}
\end{figure}

\citet{BarnesFortney2003} simulate the flux of \hd{209458~b} as if it were oblate with flattening parameter $f = 0.1$. Since they explore the dependence on a variety of transit parameters, and because they use a different method than this work, their results serve as a useful benchmark for ours. In Figures~\ref{fig:barnes-fortney-fig4} and \ref{fig:barnes-fortney-fig9}, we reproduce \citet{BarnesFortney2003} Figures~4 and 9, respectively, showing strong qualitative agreement with their findings.

\subsection{Code Performance}

\begin{figure}
    \centering
    \includegraphics[width=\linewidth]{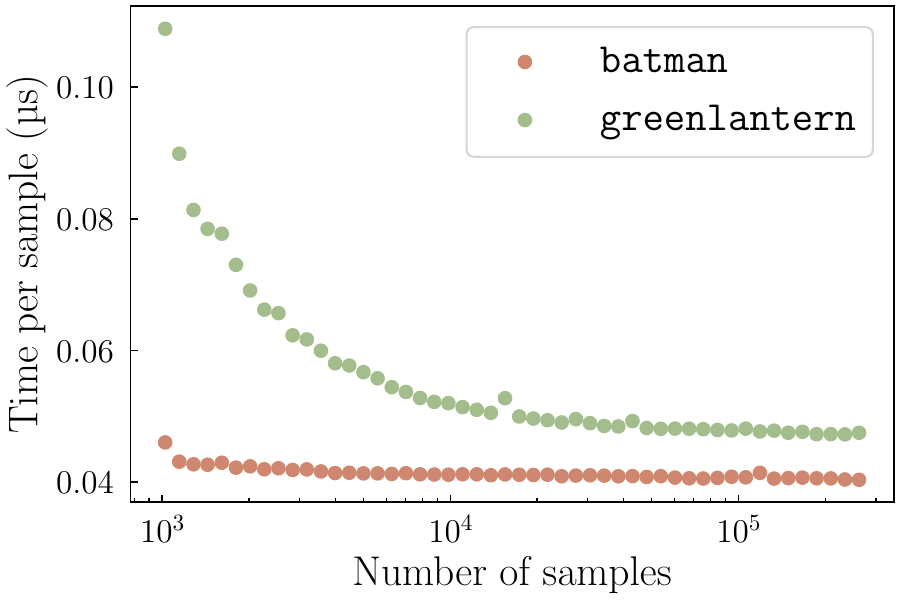}
    \caption{Performance of the code presented in this work, \texttt{greenlantern}, compared to the popular spherical transit modeling code \texttt{batman} \citep{Kreidberg2015}. By running on GPU, our code achieves similar timings per sample once at least $10^4$ samples are computed.}
    \label{fig:timing-vs-batman}
\end{figure}

\noindent The \texttt{batman} code for spherical planet transits \citep{Kreidberg2015} also provides a useful benchmark to demonstrate the speed of our approach. In Figure~\ref{fig:timing-vs-batman}, we show the wall time per flux sample for both codes, averaged over ten trials, for a variety of workload sizes. When only a few hundred points are computed, \texttt{greenlantern} is about half as fast as \texttt{batman}, but, as the workload increases to tens or hundreds of thousands of points, we asymptotically approach a comparable timing of about 0.5~\textmu s per sample. Since \texttt{greenlantern} offers additional functionality over \texttt{batman} --- the ability to compute general triaxial ellipsoid transits --- some computational cost is expected, but we show that it is not unreasonable.

\section{Data and Model Fitting}
\label{sec:data-and-fitting}

\noindent We apply our newly-developed code to the \hip{41378} system, which consists of at least five known planets, including a particularly intriguing cold puffy sub-Neptune-mass planet, \hip{41378~f} \citep{Vanderburg2016, Santerne2019}. \hip{41378~f} has an extremely low measured bulk density \citep[0.09 g/cm\(^3\);][]{Santerne2019,Harada2023} and a cold orbit far from its host star (\citealt{Becker2019,Berardo2019}; orbital period 542.08~days, \citealt{Santerne2019}). 
Previous observations with the Hubble Space Telescope have found a flat transmission spectrum shape for \hip{41378~f} \citep{Alam2022}. 

HIP 41378 was observed during K2's Campaign 5 in long-cadence mode, and during Campaign 18 in short-cadence mode. Utilizing \texttt{greenlantern}, we fit the K2 light curve data to derive a constraint on \hip{41378~f}'s rotation rate.  We used the K2 observations as reduced by \citet{Vanderburg2016} and \citet{Becker2019}. In brief, these light curves were extracted from target pixel files from a set of 20 photometric apertures and systematics corrections were performed following \citet{Vanderburg2014PASP} and \citet{Vanderburg2016ApJS}. After selecting the light curve from the aperture that minimized photometric scatter, the systematics correction was refined by a simultaneous fit along with the transits and low-frequency variability in the light curve. We flattened the light curve by dividing the resulting light curve by the best-fit low-frequency variability model. The original target pixel files for K2 Campaigns 5 and 18 are hosted by MAST: \dataset[10.17909/T9SK5H]{\doi{10.17909/T9SK5H}} and \dataset[10.17909/t9-zmte-d528]{\doi{10.17909/t9-zmte-d528}}, respectively.

The full model consists of 16 parameters, given in Table~\ref{tbl:priors} with their associated priors. The polar radius ratio ($R_a / R_\star$), equatorial radius ratio ($R_{bc} / R_\star$), and orientation angle $\eta$ are reparameterized for fitting as $r$, $\mu$, and $\nu$. The parameter $r$ is simply the geometric mean of the three ellipsoid radii; the parameterization of $\mu$ and $\nu$, which depend on flattening $f$ and orientation $\eta$, is detailed in Appendix~\ref{adx:degeneracies}. The remaining planet parameters are orbital period $P_\text{orb}$, orbital distance during transit $d / R_\star$, mid-transit time offset $t_0$, impact parameter $b$, and orientation angle $\xi$. We also require stellar limb darkening parameters $q_1$ and $q_2$, following the parameterization of \citet{Kipping2013}, and six additional parameters characterizing correlated noise. At present, we restrict our focus to quadratic limb darkening only, but extension to other limb darkening prescriptions could be an avenue for future work.

Since \hip{41378~f} has such a long orbital period, we must consider that its orbit may be eccentric. The stellar density $\rho_\star$ is constrained independently by asteroseismology \citep{Lund+2019}, which, combined with orbital period $P_\text{orb}$, sets the scaled semimajor axis $a / R_\star$ of the planet's orbit. We distinguish the orbital distance during transit with the symbol $d / R_\star$, which is equivalent to $a / R_\star$ when eccentricity $e = 0$.
This is an approximation based on the assumption that $d / R_\star$ remains constant throughout the transit, rather than varying with time.
We expect eccentricity to be small if it is nonzero, so the value of $a / R_\star$ is used to set the prior on $d / R_\star$.

We parameterize orientation as two angular parameters $\eta$ and $\xi$, as explained above; the third angle introduced in Section~\ref{sec:ellipsoid-transit} is unnecessary for an oblate ellipsoid because of the assumed symmetry. Due to degeneracies between these angles and the impact parameter $b$, we choose to explore the full range of allowed $b \in \left[-1, 1\right]$ but restrict the other parameters to the ranges $\eta \in \left[-\pi/2, \pi/2\right]$ and $\xi \in \left[0, \pi/2\right]$, rather than explore a multi-modal parameter space. $\eta$ is not fit directly but can be recovered in postprocessing; see Appendix~\ref{adx:degeneracies}.

Finally, to account for the possibility of correlated, or ``red'', noise in our dataset, we model that noise using Gaussian processes (GPs). We choose a covariance matrix that is the sum of a Mat\'ern-3/2 kernel and diagonal ``jitter'' term. The amplitudes of each covariance term and the timescale of the Mat\'ern-3/2 terms account for the final six model parameters: three for the short cadence data, and three for the long cadence data. In Table~\ref{tbl:priors}, we give the prior probabilities assumed on each of these parameters.

\begin{deluxetable*}{lcc}
    \tablecolumns{3}
    \tablehead{\colhead{Parameter} & \colhead{Symbol} & \colhead{Prior\tablenotemark{\scriptsize a}}}
    \tablecaption{Prior distributions used in Bayesian model fitting. \label{tbl:priors}}
    \startdata
        Mean radius ratio & $r$ & $\log\mathcal{U}\!\left[10^{-2}, 10^{-1}\right]$ \\
        Reparameterized\tablenotemark{\scriptsize b, c} $f$ and $\eta$ & $\mu$, $\nu$ & see Equation~\ref{eqn:mu-nu-joint-density} \\
        Orbital distance during transit\tablenotemark{\scriptsize d} & $d / R_\star$ & $\mathcal{N}\!\left[a / R_\star, 10\right]$ \\
        Orbital period (days) & $P^{}_{\text{orb}}$ & $\mathcal{N}\!\left[542.08, 0.1\right]$ \\
        Mid-transit time offset\tablenotemark{\scriptsize e} (days) & $t_0$ & $\mathcal{N}\!\left[t_{0,\text{est}}, 0.1\right]$ \\
        Impact parameter & $b$ & $\mathcal{U}\!\left[-1, 1\right]$ \\
        Orientation parameter (rad) 
        & $\xi$ & $\mathcal{U}\!\left[0, \frac{\pi}{2}\right]$ \\
        Limb darkening coefficients\tablenotemark{\scriptsize f} & $q_1, q_2$ & $\mathcal{U}\!\left[0,1\right]$ \\
        Mat\'ern-3/2 amplitude\tablenotemark{\scriptsize g, h} & $\sigma_{M,\text{lc}}$, $\sigma_{M,\text{sc}}$ & $\log\mathcal{U}\!\left[10^{-2} \sigma, 10^2 \sigma\right]$ \\
        Mat\'ern-3/2 timescale\tablenotemark{\scriptsize g, i} & $\rho_{M,\text{lc}}$, $\rho_{M,\text{sc}}$ & $\log\mathcal{U}\!\left[10^{-2} \tau, 10^2 \tau\right]$ \\
        Jitter amplitude\tablenotemark{\scriptsize g, h} & $\sigma_{J,\text{lc}}$, $\sigma_{J,\text{sc}}$ & $\log\mathcal{U}\!\left[10^{-2} \sigma, 10^2 \sigma\right]$
    \enddata
    \tablenotetext{\scriptstyle a}{We use $\mathcal{U}\!\left[a, b\right]$ to indicate the uniform distribution on the interval $\left(a, b\right)$; $\log\mathcal{U}\!\left[a, b\right]$ to indicate the log uniform, or reciprocal, distribution on the interval $\left(a, b\right)$; and $\mathcal{N}\!\left[\mu, \sigma\right]$ to indicate the normal distribution of mean $\mu$ and standard deviation $\sigma$.}
    \tablenotetext{\scriptstyle b}{Due to a model degeneracy, the parameter space of $\mu$ and $\nu$ is more easily explored by MCMC than that of $f$ and $\eta$. See Appendix~\ref{adx:degeneracies} for a full derivation.}
    \tablenotetext{\scriptstyle c}{We choose ranges of $\mu$ and $\nu$ which allow the full geometric range $f \in \left[0,1\right]$. Since the transit contains only geometric information, we choose not to place any constraints on flattening that derive from interior structure arguments \citep[as in][]{Berardo+2022}.}
    \tablenotetext{\scriptstyle d}{The measurement of stellar density $\rho_\star$ is constrained independently using asteroseismology by \citet{Lund+2019}, from which we adopt the estimate $\rho^{}_{\star,\text{est}} = 0.785$~g~cm${}^{-3}$; because the error on $\rho^{}_{\star,\text{est}}$ is small, and because the value is only used as the mean of the prior on $d / R_\star$, we treat the stellar density estimate as exact. Combined with orbital period, $\rho_{\star,\text{est}}$ uniquely sets $a / R_\star$ via $a / R_\star = \sqrt[3]{G \rho^{}_{\star,\text{est}} P_\text{orb}^2 \big/ 3 \pi}$. Fitting instead for an orbital distance during transit is a computationally efficient method for approximating nonzero eccentricity without fitting directly, assuming that the orbital distance during transit is approximately constant in time.}
    \tablenotetext{\scriptstyle e}{The mid-transit time offset mean was estimated by eye and given an artificially inflated standard deviation. This approach is intended to mitigate aliasing effects in the mid-transit time.}
    \tablenotetext{\scriptstyle f}{We adopt the quadratic limb darkening parameterization of \citet{Kipping2013}, which derives a nonlinear mapping from $q_1, q_2$ to the standard coefficients $u_1, u_2$.}
    \tablenotetext{\scriptstyle g}{We fit noise parameters for long cadence (lc) and short cadence (sc) separately.}
    \tablenotetext{\scriptstyle h}{The parameter $\sigma$ here refers to the estimated out-of-transit standard deviation of the long cadence ($\sigma_\text{lc}$) or short cadence ($\sigma_\text{sc}$) data, as appropriate.}
    \tablenotetext{\scriptstyle i}{The parameter $\tau$ here refers to the integration time for long cadence ($\tau_\text{lc}$ = 30 minutes) or short cadence ($\tau_\text{sc}$ = 1 minute) data, as appropriate.}
\end{deluxetable*}

Our fitting procedure uses \texttt{celerite} \citep{ForemanMackey+2017} to evaluate the log-likelihood of the residual between the observed data and the model, which is computed as described in Section~\ref{sec:ellipsoid-transit}. We first use \texttt{pocoMC} \citep{pocomc1,pocomc2} to efficiently explore the full parameter space with preconditioned Monte Carlo. While \texttt{pocoMC} provides several normalizing flows through the \texttt{zuko} library \citep{zuko}, we find that these may not explore near hard parameter boundaries efficiently; we instead use the neural autoregressive flow \citep[NAF,][]{Huang+2018} provided by \texttt{zuko}, as a custom flow in \texttt{pocoMC}, which has better behavior near boundaries. After this initial fitting, we use \texttt{emcee} \citep{ForemanMackey+2013} to more carefully explore around the maximum likelihood point found by \texttt{pocoMC}, using the standard deviation of the \texttt{pocoMC} samples to seed the initial walker positions.\footnote{Our final parameter posteriors are available at \dataset[10.5281/zenodo.14516951]{\doi{10.5281/zenodo.14516951}}.}

\section{Results}
\label{sec:results}

\begin{figure*}
    \centering
    \includegraphics[width=\linewidth]{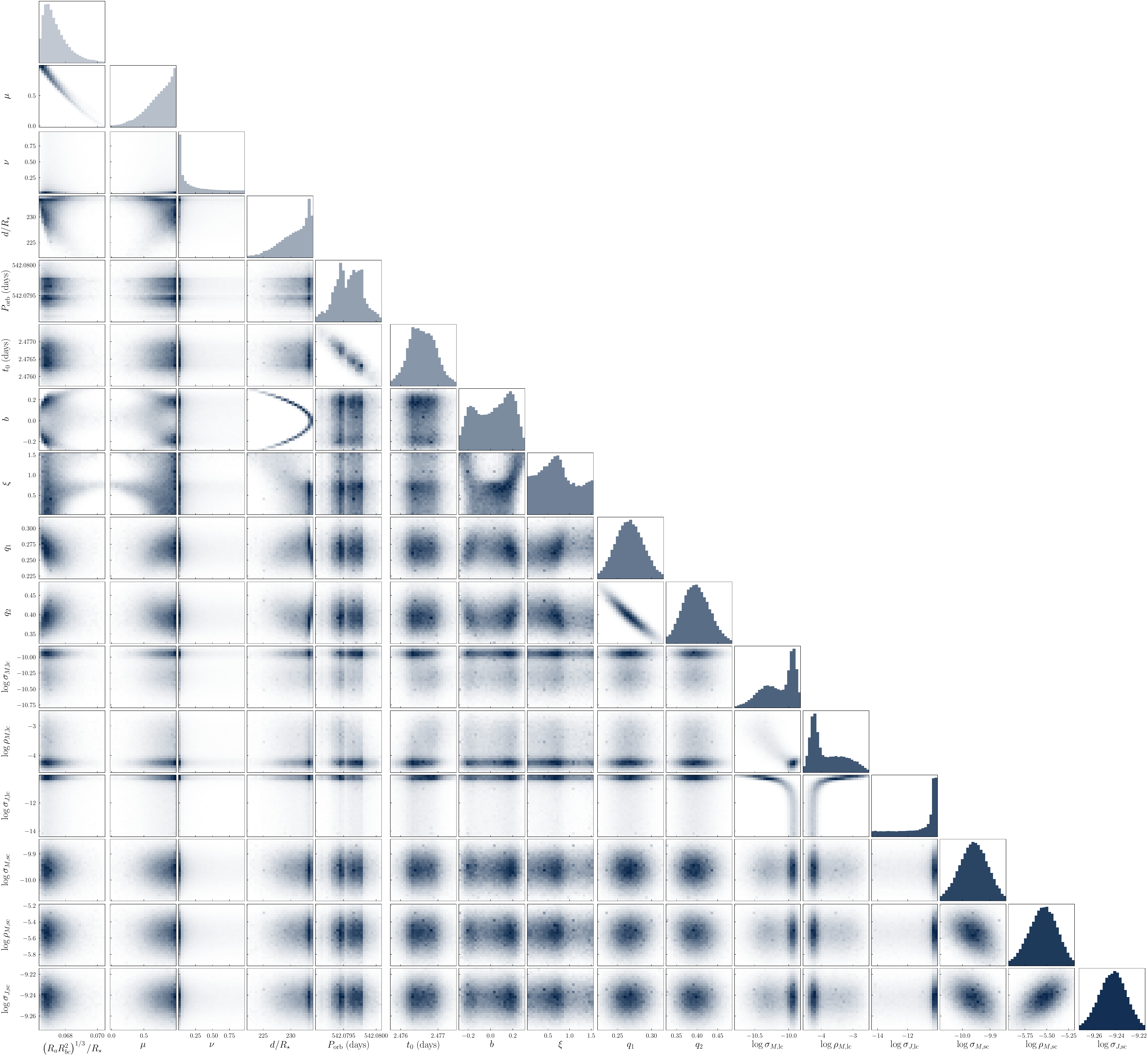}
    \caption{Joint parameter distributions when planet f is constrained to be an oblate ellipsoid with non-negative flattening.}
    \label{fig:corner-kepler}
\end{figure*}

\noindent Our MCMC fit to the transit light curve data of \hip{41378~f} yields posterior distributions for the 16 fit parameters, shown in Figure~\ref{fig:corner-kepler}. The median values, $1\sigma$ errors, and maximum likelihood values for each fitted parameter are reported in Table~\ref{tbl:fit-results}.

\begin{deluxetable*}{lccc}
    \tablecolumns{4}
    \tablehead{\colhead{Parameter} & \colhead{Symbol} & \colhead{Median and error\tablenotemark{\scriptsize a}} & \colhead{Maximum likelihood value\tablenotemark{\scriptsize b}}}
    \tablecaption{Medians, uncertainties, and maximum likelihood values of fit parameters. \label{tbl:fit-results}}
    \startdata
        Mean radius ratio & $r$ & $0.067208 \substack{+0.00097902 \\ -0.00052744}$ & $0.066599$ \\
        Reparameterized $f$ and $\eta$ & $\mu$ & $0.77319 \substack{+0.17010 \\ -0.27714}$ & $0.99271$ \\
        & $\nu$ & $0.12655 \substack{+0.49355 \\ -0.11947}$ & $0.00017187$ \\
        Orbital distance during transit & $d / R_\star$ & $231.24 \substack{+2.1496 \\ -3.6038}$ & $229.93$ \\
        Orbital period (days) & $P_{\mathrm{orb}}$ & $542.08 \substack{+0.00019790 \\ -0.00021781}$ & $542.08$ \\
        Impact parameter & $b$ & $0.041605 \substack{+0.16010 \\ -0.21322}$ & $0.18374$ \\
        Orientation parameter (rad) & $\xi$ & $0.69503 \substack{+0.55787 \\ -0.43496}$ & $1.3872$ \\
        Limb darkening coefficients & $q_1$ & $0.26744 \substack{+0.021465 \\ -0.020677}$ & $0.27007$ \\
        & $q_2$ & $0.39688 \substack{+0.035484 \\ -0.032286}$ & $0.39046$ \\
        Mat\'ern-3/2 amplitude & $\log{\sigma}_{M,\mathrm{lc}}$ & $-10.089 \substack{+0.17102 \\ -0.33020}$ & $-10.287$ \\
        & $\log{\sigma}_{M,\mathrm{sc}}$ & $-9.9589 \substack{+0.051082 \\ -0.052075}$ & $-9.9807$ \\
        Mat\'ern-3/2 timescale & $\log{\rho}_{M,\mathrm{lc}}$ & $-4.0116 \substack{+0.85785 \\ -0.29395}$ & $-3.6581$ \\
        & $\log{\rho}_{M,\mathrm{sc}}$ & $-5.5383 \substack{+0.15865 \\ -0.17078}$ & $-5.4778$ \\
        Jitter amplitude & $\log{\sigma}_{J,\mathrm{lc}}$ & $-10.524 \substack{+0.37890 \\ -2.5112}$ & $-10.271$ \\
        & $\log{\sigma}_{J,\mathrm{sc}}$ & $-9.2430 \substack{+0.012290 \\ -0.012897}$ & $-9.2393$ \\
    \enddata
    \tablenotetext{\scriptstyle a}{We report the 50\% quantile as the median; the error is computed from the 15.85\% and 84.15\% quantiles, such that the quoted interval contains approximately 68.3\% of the probability density.}
    \tablenotetext{\scriptstyle b}{Since some posterior distributions are bimodal, we additionally provide the value of each parameter at the maximum likelihood sampled by the MCMC.}
\end{deluxetable*}

\begin{figure}
    \centering
    \includegraphics[width=\linewidth]{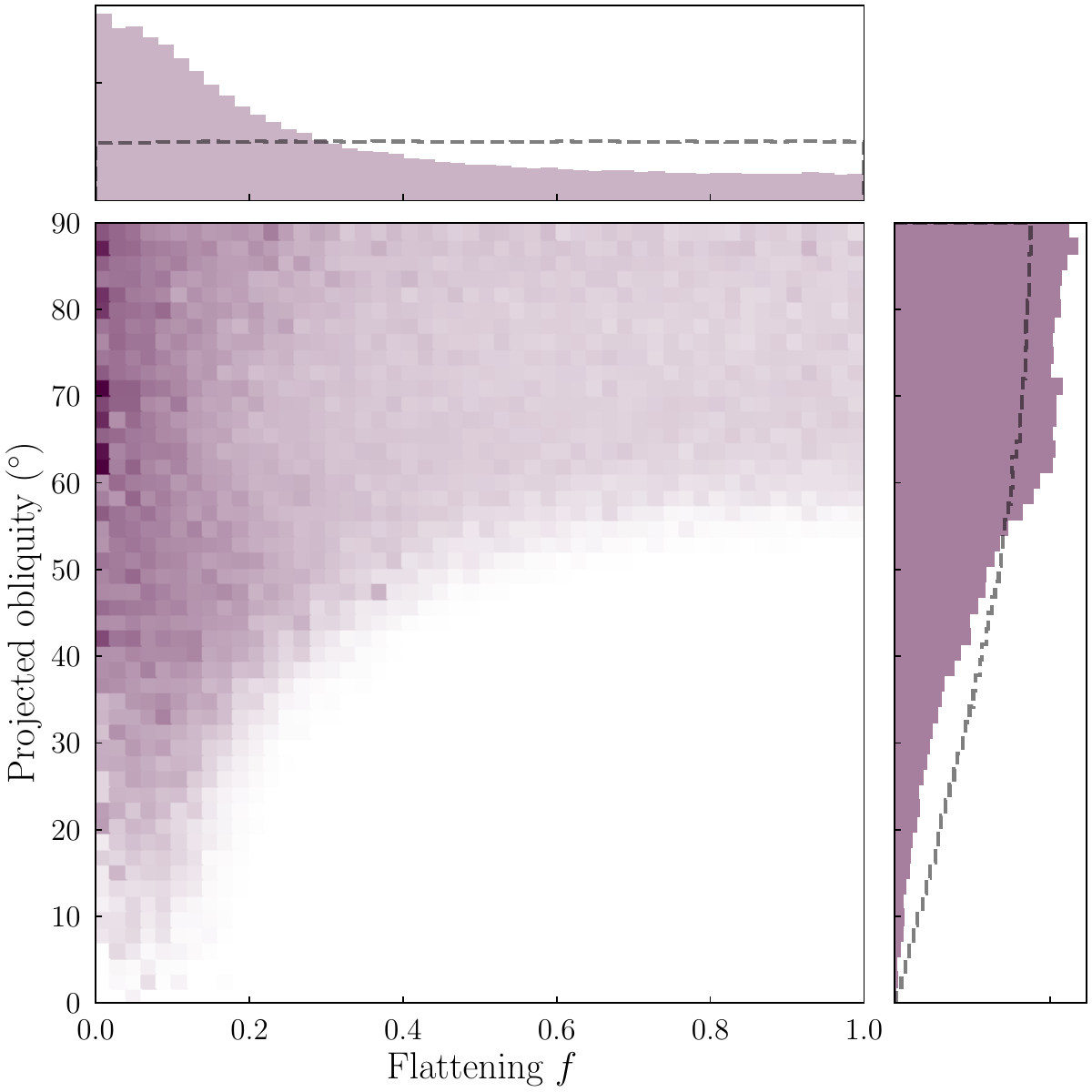}
    \caption{Joint distribution of flattening $f$ and projected obliquity, which we define as the angle between the $\hat{\bm{x}}$ and spin angular momentum vectors, for \hip{41378~f}, derived from the transit light curve fitting using the model described in Section \ref{sec:ellipsoid-transit}. The central panel displays the density of solutions in the joint parameter space, with darker shades indicating a higher density of allowed solutions. The marginal histograms above and to the right of the central panel show the projected distributions of $f$ and projected obliquity, respectively. We overlay the prior probability distributions on the marginalized flattening and obliquity for comparison, using dashed gray lines. The posterior distributions highlight the range of possible planetary shapes (flattening $f$; lower values are more strongly preferred) and axial tilts (projected obliquity; all values are technically allowed by the model). The results suggest a wide range of possible obliquities, indicating a range of potential dynamical histories and orientations of \hip{41378~f}.}
    \label{fig:joint-flattening-obliquity}
\end{figure}

In Figure \ref{fig:joint-flattening-obliquity}, we present the joint distribution of flattening $f$ and projected obliquity for \hip{41378~f}. Projected obliquity here is defined as the true angle between the $\hat{\bm{x}}$ and spin angular momentum vectors.
This constraint on planetary flattening provides an opportunity to constrain the planetary rotation period of \hip{41378~f}. 
The rotation period of an exoplanet $P_\text{rot}$ and its flattening $f$ are related by
\begin{equation}
    P_\text{rot} = 2\pi \sqrt{\frac{R_\text{eq}^3}{G M_p \left(2 f - 3 J_2\right)}},
    \label{eq:prot}
\end{equation}
where $G$ is the Newtonian gravitational constant, $R_\text{eq}$ is the equatorial radius, $M_\text{p}$ is the mass, and $J_2$ is the second spherical mass moment \citep{Hubbard1984,CarterWinn2010a}. $J_2$ and $f$ are not independent, since a flattened planet can have a different interior structure than its spherical counterpart. 
For a body with no mass asymmetry, $J_2 = 0$, while more oblate bodies have larger $J_2$ values: for example, Jupiter has a $J_2 = 0.15$. 
$J_2$ depends on the unknown interior structure of \hip{41378~f}, so a direct measurement of $P_\text{rot}$ is not possible even with the constraints we have derived on \hip{41378~f}'s flattening $f$. 
However, as inspection of Equation~\ref{eq:prot} will reveal, the effect of a larger $J_2$ moment is to increase the inferred planetary rotational period. 
As such, the rotational period calculated using Equation~\ref{eq:prot} under the assumption $J_2 = 0$ \citep[as done in][]{Seager2002} will provide a lower limit on the planetary rotational period. 
Using our derived posterior on $f$ and random draws from normal distributions of $R_\star$ and $M_\text{p}$ with parameters from \citet{Santerne2019}, we compute a distribution of computed rotation periods for \hip{41378~f}, under the assumption $J_2 = 0$.
We place a lower limit $P_\text{rot} \geq \protval$~hr on the rotation period of \hip{41378~f}, as well as an upper limit of $f \leq \flatval$ on the planet's flattening, both at the 95\% confidence level.

We can compare this limit on the rotation period to one calculated using the \cite{BarnesFortney2003} formulation of the Darwin-Radau relation (Equation 5 of that paper), which parameterizes the planetary interior structure into a $\mathds{C}$ parameter describing the ratio of the planet's moment of inertia to $M^{}_\text{p} R_\text{eq}^2$. For gas giants, this parameter is inexact, but can be approximated as $\mathds{C} = 0.25$ \citep{Hubbard1984}. This formulation gives a lower limit on the rotation period of $P_\text{rot} \geq \protvaldr$~hr.

\begin{figure*}
    \centering
    \includegraphics[width=\linewidth]{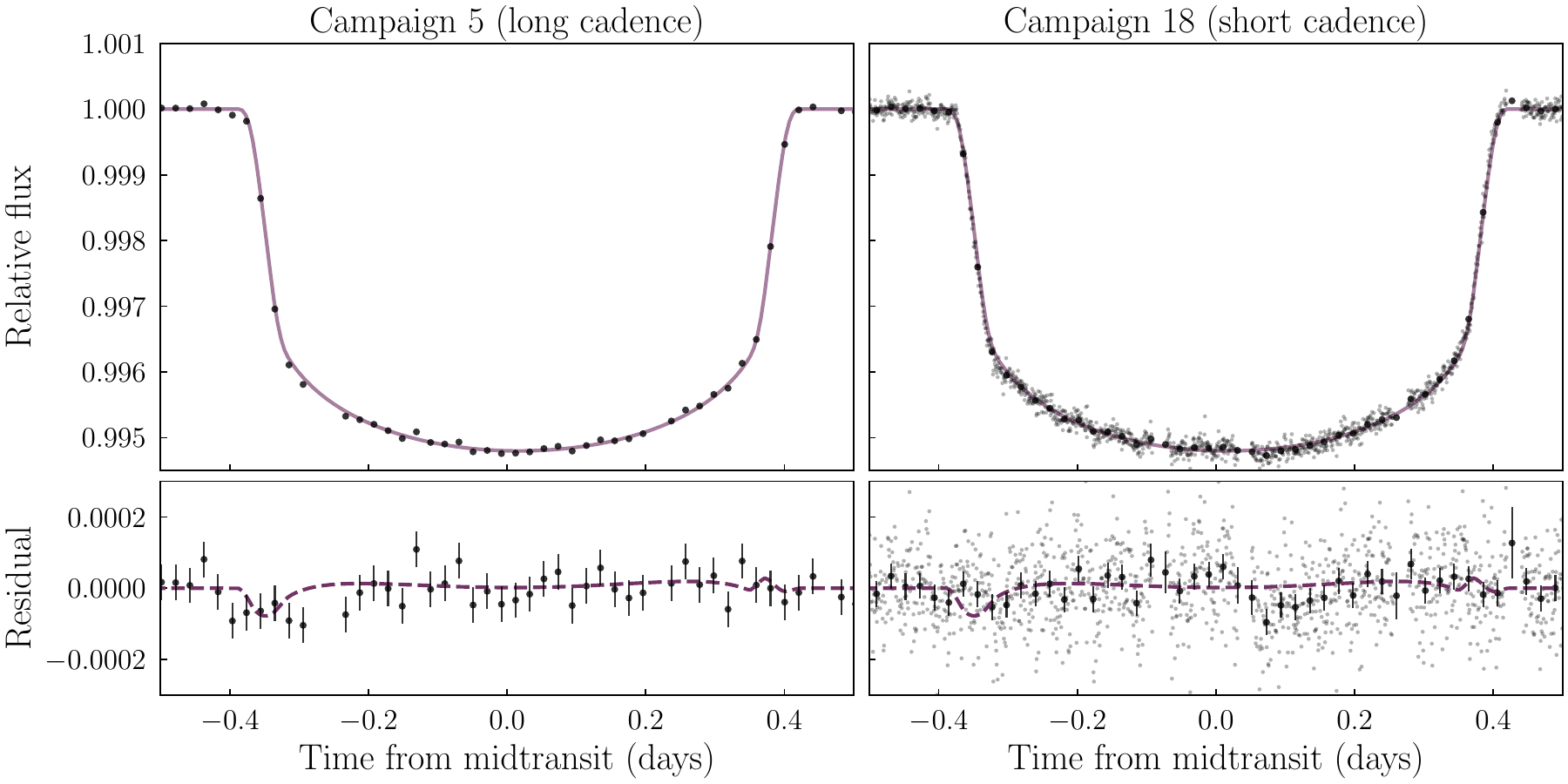}
    \caption{Long- and short-cadence transit light curve data for \hip{41378~f}, folded with orbital period $P_\text{orb} = 542.08$~days. The reduced long-cadence data is shown in the left panels in scattered points; the short-cadence data is shown on the right with small points, then binned to 30~minutes and plotted with large points. The solid purple curve shows a randomly-drawn model from the posterior for comparison. The lower row presents the residual between the maximum likelihood model and the data. If we select the maximum likelihood model with $f > 0.1$, we obtain the residual shown in the dashed purple curve, which is disfavored with log-likelihood difference $\Delta \left(\log{\mathcal{L}}\right) \approx -3.4$}
    \label{fig:data-and-solution}
\end{figure*}

In Figure~\ref{fig:data-and-solution}, we show the maximum likelihood model from the posterior distribution and a the residuals of the maximum likelihood model constrained to $f > 0.1$. The more spherical model is favored over the flattened one.

\section{Discussion}
\label{sec:discussion}

\noindent In this paper, we introduce \texttt{greenlantern}: a new GPU-accelerated code designed to fit transit light curves of ellipsoidal planets. This code demonstrates high accuracy when compared to existing models that assume spherical planets, and it maintains competitive time performance, especially for use cases requiring large numbers of samples.

We then apply this code to the cold super-puff \hip{41378~f} and establish a lower limit to its rotation period of $P_\text{rot} \geq \protval$ hours. 
Rotation rates have previously been measured or constrained for a small number of directly imaged planets through spectroscopy \citep{Wang2021, Parker2024, Morris2024} and a small number of massive short-period planets via transit data \citep{CarterWinn2010a, Zhu2014, Biersteker2017}.
The ability to constrain planetary rotation and obliquity with this new code, particularly for lower-mass planets, will allow a new dimension of constraints on the processes of planet formation.  

\subsection{Implications for Formation}

\noindent This work reveals a broad range of possible obliquities for \hip{41378~f} (Figure \ref{fig:joint-flattening-obliquity}). 
While giant planets can attain primordial planetary obliquities via interactions with the circumplanetary disk \citep{Martin2021} or disk fragmentation \citep{Jennings2021}, \hip{41378~f}'s low mass \citep[$12 \pm 3 M_{\oplus}$;][]{Santerne2019} precludes both of these mechanisms.
For planets of this mass, large obliquities could plausibly arise from dynamical interactions between planets, including spin-orbit resonances \citep{Ward2004, LiGongie2021, Millholland2019a, Saillenfest2019, Millholland2024}, interactions between planets and their satellites \citep{Saillenfest2021b, Saillenfest2022}, interactions between the protoplanetary disk and a young planet \citep{Millholland2019b, Su2020}, or collisions between the planet and another object in the system \citep{Slattery1992, Morbidelli2012}. 
Confirming a significant obliquity for \hip{41378~f} would be compelling, especially since this planet has no known nearby companion planets to drive strong planet-planet interactions \citep[its nearest known companion planet, HIP~41378~e, has an approxmiate period ratio of $P_{\mathrm{orb,f}} / P_{\mathrm{orb,e}} \approx 1.46$,][]{Santerne2019}.
For long-period planets like \hip{41378~f}, large obliquities or rapid rotation rates offer important clues about past planet-planet interactions and the historical spacing within planetary systems, as discussed by \citet{Li2020}. The processes driving the rotational evolution of cold planets are markedly different from those affecting hot Jupiters, which often undergo dynamical erasure of their histories via tidal interactions and therefore offer less insight towards their formation histories.
\citet{Lu2024} suggests that a high planetary obliquity for \hip{41378~f} could indicate that \hip{41378~f} might have been part of a resonant chain and was captured into a spin-orbit resonance during convergent migration.

Based on the rotation period limit for \hip{41378~f} derived in this work, it is likely that \hip{41378~f} is a less rapid rotator than the Solar System gas giants. While the quality of the current light curve is insufficient to securely measure \hip{41378~f}'s obliquity, improved constraints with future data would allow a more direct test of the importance of past impacts or planet-disk interactions in setting its current dynamical state, including a direct test of the hypothesis of \citet{Lu2024}. 

\subsection{Possibility of Rings}

\noindent \hip{41378~f} has a Jupiter-like radius \citep[$9.2 \pm 0.1 R_{\oplus}$;][]{Vanderburg2016}, sub-Neptune mass \citep[$12 \pm 3 M_{\oplus}$;][]{Santerne2019}, and an anomalously low density (0.09 g cm$^{-3}$). Combined with \hip{41378~f}'s low effective temperature \citep[$\sim294$ K;][]{Santerne2019} and the old age of its host star \citep[$\sim2$ Gyr;][]{Lund+2019}, the planet's structure subverts expectations. Its range of allowable core masses are less massive than expected for a planet of this size \citep{Belkovski2022}, resulting in a theoretical challenge of how the planet attained its observed mass and density. 

One explanation for its large radius could be the presence of circumplanetary rings \citep{Akinsanmi2020, Piro2020}, possibly due to migrating and disintegrating exomoons \citep{Saillenfest2023}, which would increase the apparent planetary radius. If viewed in a face-on geometry, these rings could mimic the shape of a ringless planet. 
HST transmission spectra of this planet are consistent with the rings hypothesis \citep{Alam2022}, which would be expected to cause a flat transmission spectrum \citep{Ohno2022}. 

The results of this paper demonstrate flattening of $f \leq \flatval$ is consistent with the shape of the transit light curve. In Section \ref{sec:results}, we discuss how this constrained amount of flattening can be used to infer the planetary rotation rate, assuming that rotational deformation is the primary cause of non-sphericity.
However, it is important to note that if \hip{41378~f} does host circumplanetary rings, the observed flattening would not be due to the planet's rotation, but rather the geometry of the rings. Consequently, in that case, our stated constraints on the planet's rotation would no longer apply.

\subsection{Improving Constraint with JWST}

\noindent As demonstrated in Figure \ref{fig:barnes-fortney-fig9}, the signal that allows the detection of flattening in a light curve is small in amplitude and occurs at transit ingress and egress only. As a result, such measurements can only be made with sufficiently high-precision and high-cadence data. In this work, we use K2 data on a long-period planet; however, this data quality provides only an upper limit on flattening. 
To improve upon the constraint of this work, we need a high precision light curve to precisely characterize the exact shape of ingress and egress events. 
\citet{Liu2024} demonstrates that for a Saturn-like oblateness, a greater-than-Earth-like obliquity ($>20^\circ$), and a Jupiter-like planet, one transit of JWST data could recover the flattening. 
\begin{deluxetable}{lcc}
    \tablecolumns{3}
    \tablehead{\colhead{Instrument mode} & \colhead{Cadence} & \colhead{Noise estimate (ppm)}}
    \tablecaption{Estimated noise of follow-up JWST observations. \label{tbl:jwst-noise}}
    \startdata
        NIRISS, substrip 256 & 1~min & 21.71 \\
        NIRISS, substrip 256 & 2~min & 15.35 \\
        NIRISS, substrip 96 & 1~min & 21.71 \\
        NIRISS, substrip 96 & 2~min & 15.35 \\
    \enddata
\end{deluxetable}

To evaluate the improvements on the flattening constraint that would be possible for \hip{41378~f} with JWST data, we simulated the expected noise for observing the transit of \hip{41378~f} using \texttt{PandExo} \citep{Batalha+2017} for NIRISS. We focus our noise estimates on NIRISS since it has already been shown to produce high-precision light curves and achieves 5~ppm RMS for 1-hour bins when observing WASP-18~b \citep{Coulombe+2023}. In our \texttt{PandExo} simulation, we use a PHOENIX stellar model grid for the stellar properties of \hip{41378} ($T = 6226$~K, $\left[\text{Fe}/\text{H}\right] = -0.11$, $\log{g} = 7.982$). We generate our own custom planet model using \texttt{batman} \citep{Kreidberg2015} and the planet parameter estimates from \citet{Berardo2019}. We input this photometric light curve into \texttt{PandExo}. We set the saturation limit to 80\% and compute noise estimates for 2- and 1-minute cadence for one transit, since only one will occur during cycle 4. We find that the 2-minute cadence observations would be expected to produce 15.35~ppm of noise and the 1-minute cadence observations would produce 21.71~ppm of noise.

\begin{figure}
    \centering
    \includegraphics[width=\linewidth]{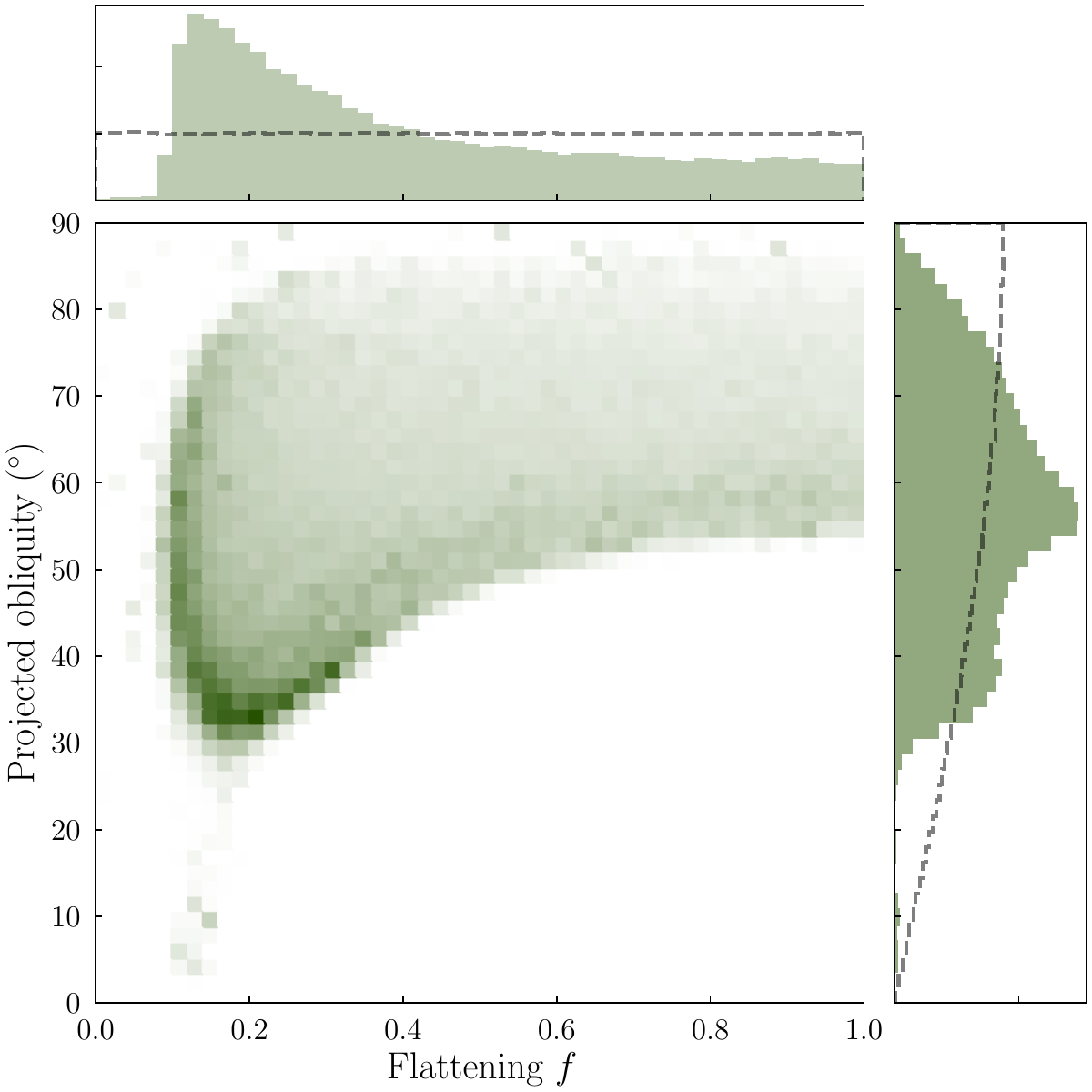}
    \caption{Joint posterior of flattening and obliquity for a simulated light curve of a flattened planet with $f = 0.15$. For this test case, fitting with \texttt{greenlantern} excludes the $f = 0$ model. As in Figure~\ref{fig:joint-flattening-obliquity}, the prior probability distribution for the marginalized quantities is shown with dashed gray lines.}
    \label{fig:sim-joint-flattening-obliquity}
\end{figure}

Based on these noise estimates, we simulate a transit light curve for \hip{41378~f} with 20~ppm white noise and no red noise using a 2-minute cadence. We adopt similar parameter values as those from Table~\ref{tbl:fit-results} but inject test values for flattening ($f = 0.15$) and orientation ($\eta = 0^\circ$ and $\xi \approx 29^\circ$).
We find that, if these were the true values, photometric data from JWST could definitively exclude a spherical planet shape, $f = 0$. The joint posterior of flattening and obliquity is shown in Figure~\ref{fig:sim-joint-flattening-obliquity}.

\section{Conclusions}
\label{sec:conclusions}

\noindent In this paper, we present a constraint on the rotation period of a sub-Neptune-mass exoplanet, \hip{41378~f}, using K2 transit light curve data. 
By developing and applying a new GPU-accelerated code, \texttt{greenlantern}, which models the transits of ellipsoidal planets, we are able to derive posterior distributions for the planet's flattening and obliquity. These constraints allow us to place a lower limit on \hip{41378~f}'s rotation period of $P_\text{rot} \geq \protval$ hours at the 95\% confidence level, suggesting a slower rotation rate than is seen for the Solar System gas giants.
Future observations, especially with high-precision instruments like JWST, will allow for even tighter constraints on planetary deformation and rotation rate, further enhancing our understanding of exoplanetary formation and evolution processes.

\textit{Note:} During the late stages of manuscript preparation, we became aware of two other works on similar topics. \citet{Cassese2024arXiv} described the development of another GPU-based code for modeling oblate planets, while \citet{LammersWinn2024} reported on similar constraints on the rotation period of a different super-puff planet, Kepler 51 d. During the referee process, we became aware of additional work, \citet{Dholakia2024}, which provides a JAX-based method to fit oblate planets and finds a spin period constraint for the hot inflated Neptune WASP-107 b. These works are complementary to ours and highlight the growing importance of oblateness measurements in the era of precise photometry from JWST.

\section*{Acknowledgments}
This research has made use of NASA's Astrophysics Data System; the NASA Exoplanet Archive, which is operated by the California Institute of Technology, under contract with the National Aeronautics and Space Administration under the Exoplanet Exploration Program; and the SIMBAD database, operated at CDS, Strasbourg, France. The National Geographic Society--Palomar Observatory Sky Atlas (POSS-I) was made by the California Institute of Technology with grants from the National Geographic Society. The Oschin Schmidt Telescope is operated by the California Institute of Technology and Palomar Observatory. This paper includes data collected by the \textit{Kepler}/K2 mission. Funding for the \textit{Kepler}/K2 mission was provided by the NASA Science Mission directorate.

EMP gratefully acknowledges funding from the Heising-Simons Foundation through their 51 Pegasi b Postdoctoral Fellowship, as well as helpful conversation with Shashank Dholakia (University of Queensland) and Prof. Joshua S. Speagle (University of Toronto). Z.L.D. would like to thank the generous support of the MIT Presidential Fellowship, the MIT Collamore-Rogers Fellowship, the MIT Teaching Development Fellowship, and to acknowledge that this material is based upon work supported by the National Science Foundation Graduate Research Fellowship under Grant No. 1745302. LAR gratefully acknowledges support from the Research Corporation for Science Advancement through a Cottrell Scholar Award. This material is based upon work supported by the National Aeronautics and Space Administration under Grant No. 80NSSC25K7159 issued through the Exoplanets Research Program (XRP).

\facility{\textit{Kepler}/K2, MAST, ADS}

\software{\texttt{batman} \citep{Kreidberg2015}, \texttt{celerite} \citep{ForemanMackey+2017}, \texttt{emcee} \citep{ForemanMackey+2013}, \texttt{PandExo} \citep{Batalha+2017}, \texttt{pocoMC} \citep{pocomc2}}

\appendix
\section{Model Degeneracies}
\label{adx:degeneracies}

\noindent In the case of the oblate spheroid model we apply to \hip{41378~f}, we take two of the ellipsoid axes to be equal, $b = c$. Evaluating Equations \ref{eqn:dx}, \ref{eqn:dy1}, and \ref{eqn:dy2} with this constraint reveals a subtle degeneracy between parameters: For a given value of flattening $f$ and orientation angle $\eta$, we compute a value
\begin{equation}
    \mu\!\left(f, \eta\right) = \left(f - 1\right)^2 + f \left(f - 2\right) \cos{2\eta}.
    \label{eqn:mu}
\end{equation}
For all other parameters (mean radius ratio, orbital distance during transit, orbital period, mid-transit time offset, impact parameter, orientation angle $\xi$, and limb darkening) fixed, any model with $\mu_1 = \mu\!\left(f_1, \eta_1\right)$ is indistinguishable from another with $\mu_2 = \mu\!\left(f_2, \eta_2\right)$, provided $\mu_1 = \mu_2$, even when $f_1 \neq f_2$ and $\eta_1 \neq \eta_2$; Figure~\ref{fig:degeneracy} provides a simple illustration of this effect. We show several contours of fixed $\mu$ in the parameter space of $f$ and $\eta$ in Figure~\ref{fig:mu-contours}.

If we were uninterested in constraining the rotation period of the planet, we could use a projected flattening to avoid this degeneracy; however, the rotation rate depends on the true flattening, and so a more careful treatment is needed. The space of $\left(f, \eta\right)$ is difficult to traverse with MCMC, so we develop the following reparameterization. The minimum allowed value of $f$ for a given $\mu$ is given by
\begin{equation}
    f_\text{min} = 1 - \sqrt{\frac{\mu + 1}{2}}.
\end{equation}
Developing an invertible mapping requires two parameters, so we introduce a new quantity $\nu \in \left[0, 1\right]$ such that
\begin{equation}
    f = f_\text{min} + \left(1 - f_\text{min}\right) \nu.
\end{equation}
Solving for $\nu$, we find
\begin{widetext}
\begin{equation}
    \nu\!\left(f, \eta\right) = \frac{2 + 2 f \left(f - 2\right) \cos^2{\eta} + \left(f - 1\right) \sqrt{4 + 2 f \left(f - 2\right) \left(1 + \cos{2 \eta}\right)}}{2 + f \left(f - 2\right) \left(1 + \cos{2\eta}\right)}.
\end{equation}
\end{widetext}
The determinant of the Jacobian $J$ of the mapping from $\left(f, \eta\right)$ to $\left(\mu, \nu\right)$ is
\begin{equation}
    \det J = \frac{2^{3/2} f \left(f - 2\right) \sin{2 \eta}}{\sqrt{2 + f \left(f - 2\right) \left(1 + \cos{2\eta}\right)}}.
\end{equation}
If we choose the joint density
\begin{widetext}
\begin{equation}
    p\!\left(\mu, \nu\right) = 2^{-3/2} \left[\nu \left(\nu - 2\right) \left(\mu \left(\nu - 1\right)^2 + \nu \left(\nu - 2\right) - 1\right)\right]^{-1/2},
    \label{eqn:mu-nu-joint-density}
\end{equation}
\end{widetext}
then we recover the target joint density
\begin{equation}
    p\!\left(f, \eta\right) \propto \cos{\eta},
\end{equation}
the product of a uniform prior on $f$ and a cosine prior on $\eta$. The prior term $\cos{\eta}$ is determined by the explicit form of the Haar metric and ensures that the rotations generated are uniformly distributed on the sphere.

Sampling directly from the distribution of $\left(\mu, \nu\right)$ can be accomplished using inverse transform sampling. We draw two independent uniform random variates $\left(u, w\right) \in \left[0, 1\right]$. The marginalized distribution of $\nu$ is given by
\begin{equation}
    p\!\left(\nu\right) = \int\limits_{-1}^1 p\!\left(\mu, \nu\right) \diff \mu = \frac{1}{\nu \left(2 - \nu\right) + \sqrt{\nu \left(2 - \nu\right)}}
\end{equation}
and its cumulative distribution is
\begin{equation}
    F\!\left(\nu\right) = \int\limits_0^\nu p\!\left(\nu^\prime\right) \diff \nu^\prime = \frac{\nu - \sqrt{\nu \left(2 - \nu\right)}}{\nu - 1}.
\end{equation}
Inverting $F\!\left(\nu\right) = u$ leads to
\begin{equation}
    \nu = \frac{u^2}{u^2 - 2 u + 2}.
\end{equation}
To randomly draw $\mu$, given this choice for $\nu$, we require its conditional cumulative distribution,
\begin{widetext}
\begin{equation}
    F\!\left(\mu | \nu\right) = \int\limits_{-1}^\mu \frac{p\!\left(\mu^\prime, \nu\right)}{p\!\left(\nu\right)} \diff \mu^\prime = \frac{\left(1 + \sqrt{\nu \left(2 - \nu\right)}\right) \left(2 - \sqrt{2 - 2 \mu \left(\nu - 1\right)^2 - 2 \nu \left(\nu - 2\right)}\right)}{2 \left(\nu - 1\right)^2}.
\end{equation}
We solve $F\!\left(\mu | \nu\right) = w$ for $\mu$ and find
\begin{equation}
    \mu = \frac{4 w + 4 \left(w^2 - w\right) \sqrt{\nu \left(2 - \nu\right)} + 2 \left(\nu \left(\nu - 2\right) - 1\right) w^2 - \left(\nu - 1\right)^2 }{\left(\nu - 1\right)^2}.
\end{equation}
\end{widetext}

\begin{figure}
    \centering
    \resizebox{\linewidth}{!}{%
        \begin{tikzpicture}[scale=1.75]%
            \begin{scope}[shift={(0,0)}]%
                \node[anchor=west] (a) at (-1.1,0.7) {(a) Front view (indistinguishable)};%
                \draw[line width=2pt, purple] (0,0) ellipse[x radius=1, y radius=0.5] node[black] (e1) {$\mu = -\frac{1}{2}$};%
                \draw[line width=2pt, purple] (2.25,0) ellipse[x radius=1, y radius=0.5] node[black] (e2) {$\mu = -\frac{1}{2}$};%
            \end{scope}%
            \begin{scope}[shift={(0,-1.5)}]%
                \node[anchor=west] (a) at (-1.1,0.7) {(b) Side view (never visible)};%
                \draw[line width=2pt, purple] (0,0) ellipse[x radius=1, y radius=0.5] node[black, align=left] (e3) {$f = \frac{1}{2}$ \\ $\eta = 0^\circ$};%
                \draw[line width=2pt, purple] (2.25,0) ellipse[x radius=1, y radius=0.333, rotate=-23.3] node[black, align=left] (e4) {$f = \frac{2}{3}$ \\ $\eta \approx 23^\circ$};%
            \end{scope}%
        \end{tikzpicture}%
    }
    \caption{A simple illustration of the model degeneracy present when the two semiaxes $b$ and $c$ are equal and the planet is not tidally locked to its host star. For a given value of $\mu$, there is a range of flattenings $f$ and orientation angles $\eta$ that can produce an identical projection. We show here that the same projection (upper panel) can be produced by two different geometries (lower panel).}
    \label{fig:degeneracy}
\end{figure}

\begin{figure}
    \centering
    \includegraphics[width=\linewidth]{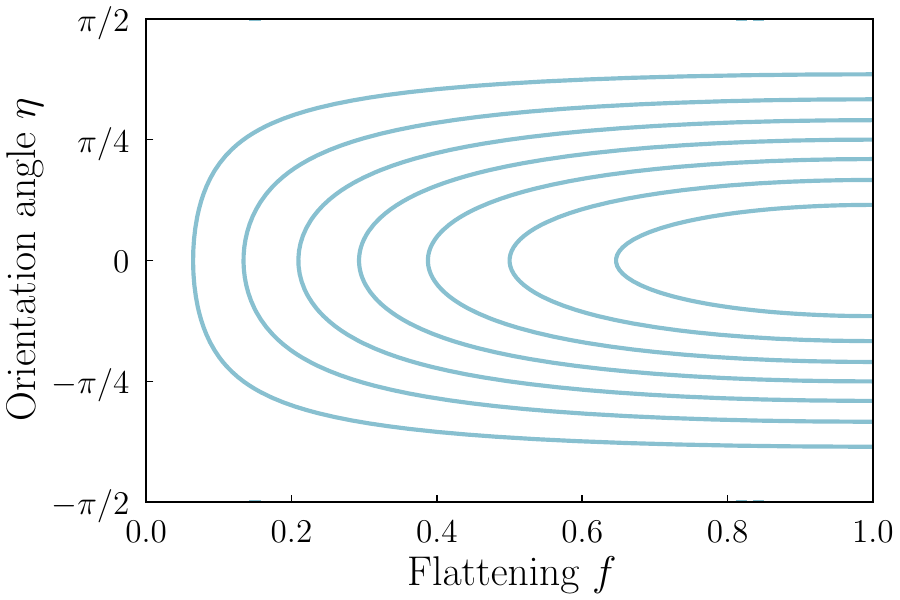}
    \caption{Contours of constant $\mu$, as defined in Equation~\ref{eqn:mu}. Along these contours, models are indistinguishable, and the Bayesian parameter posteriors are influenced only by the prior.}
    \label{fig:mu-contours}
\end{figure}

\section{The Effect of Tidal Locking}

\noindent Planets with long orbital periods, such as \hip{41378~f}, are very unlikely to be tidally locked, and we have made this assumption in the current manuscript. On the other hand, planets on short orbital periods may experience significant tidal distortion and become tidally locked to their host stars. In Figure~\ref{fig:tidally-locked}, we illustrate these two limiting cases. 

We can address the case of a tidally-locked planet by using the alternative coordinate mapping
\begin{equation}
    \begin{pmatrix} x^\prime \\ y^\prime \\ z^\prime \end{pmatrix} = R\!\left(\zeta, \eta, \xi\right) \left[R_x\!\left(-\alpha\right) R_y\!\left(\beta\right) \begin{pmatrix} x \\ y \\ z \end{pmatrix} + \begin{pmatrix} 0 \\ 0 \\ d \end{pmatrix}\right].
\end{equation}
Intuitively, this can be understood as rotating the point $\left(x, y, z\right)$ to align with the $\bm{\hat{z}}$-axis, rather than rotating the $\bm{\hat{z}}$-axis (see Equation~\ref{eqn:coord-map}). We find
\begin{align}
    \delta x = &\left[a^2 \left(\sin{\beta} \left(r_3 \cos{\alpha} + r_2 \sin{\alpha}\right) - r_1 \cos{\beta}\right)^2\right. + \nonumber\\
    &\hphantom{\left[\right.} \left.b^2 \left(\sin{\beta} \left(r_6 \cos{\alpha} + r_5 \sin{\alpha}\right) - r_4 \cos{\beta}\right)^2\right. + \nonumber\\
    &\hphantom{\left[\right.} \left.c^2 \left(\sin{\beta} \left(r_9 \cos{\alpha} + r_8 \sin{\alpha}\right) - r_7 \cos{\beta}\right)^2 \right]^{1/2},
    \label{eqn:dx-locked}
\end{align}
\begin{widetext}
\begin{align}
    \delta y_1 = &\left[a^2 \left(\sin{\beta} \left(r_3 \cos{\alpha} + r_2 \sin{\alpha}\right) - r_1 \cos{\beta}\right) \left(r_3 \sin{\alpha} - r_2 \cos{\alpha}\right)\right. + \nonumber\\
    &\hphantom{\left[\right.} \left.b^2 \left(\sin{\beta} \left(r_6 \cos{\alpha} + r_5 \sin{\alpha}\right) - r_4 \cos{\beta}\right) \left(r_6 \sin{\alpha} - r_5 \cos{\alpha}\right)\right. + \nonumber\\
    &\hphantom{\left[\right.} \left.c^2 \left(\sin{\beta} \left(r_9 \cos{\alpha} + r_8 \sin{\alpha}\right) - r_7 \cos{\beta}\right) \left(r_9 \sin{\alpha} - r_8 \cos{\alpha}\right)\right] \bigg/ \delta x,
    \label{eqn:dy1-locked}
\end{align}
and
\begin{align}
    \delta y_2 = &\left[a^2 b^2 \left(r_7 \sin{\beta} + \cos{\beta} \left(r_9 \cos{\alpha} + r_8 \sin{\alpha}\right)\right)^2\right. + \nonumber\\
    &\hphantom{\left[\right.} \left.a^2 c^2 \left(r_4 \sin{\beta} + \cos{\beta} \left(r_6 \cos{\alpha} + r_5 \sin{\alpha}\right)\right)^2\right. + \nonumber\\
    &\hphantom{\left[\right.} \left.b^2 c^2 \left(r_1 \sin{\beta} + \cos{\beta} \left(r_3 \cos{\alpha} + r_2 \sin{\alpha}\right)\right)^2\right]^{1/2} \bigg/ \delta x.
    \label{eqn:dy2-locked}
\end{align}
\end{widetext}
By replacing Equations \ref{eqn:dx}, \ref{eqn:dy1} and \ref{eqn:dy2} with Equations \ref{eqn:dx-locked}, \ref{eqn:dy1-locked}, and \ref{eqn:dy2-locked}, we can simulate light curves of tidally locked planets using the same method presented in Section~\ref{sec:ellipsoid-transit}.

\begin{figure}
    \centering
    \resizebox{\linewidth}{!}{%
        \begin{tikzpicture}[scale=2]%
            \draw[line width=3pt, yellow] (0,0) ++(-180:1) arc (-180:0:1);%
            \draw[line width=2pt, purple] (0,0) ++(-30:2) ellipse[x radius=0.2, y radius=0.4];
            \draw[line width=2pt, purple] (0,0) ++(-90:2) ellipse[x radius=0.2, y radius=0.4];
            \draw[line width=2pt, purple] (0,0) ++(-150:2) ellipse[x radius=0.2, y radius=0.4];
            \draw[line width=2pt, purple!70!black, dashed] (0,0) ++(-30:2) ellipse[x radius=0.2, y radius=0.4, rotate=60];
            \draw[line width=2pt, purple!70!black, dashed] (0,0) ++(-90:2) ellipse[x radius=0.2, y radius=0.4];
            \draw[line width=2pt, purple!70!black, dashed] (0,0) ++(-150:2) ellipse[x radius=0.2, y radius=0.4, rotate=-60];
            \draw[line width=1pt, black, dashed] (0,0) ++(-180:2) arc (-180:0:2);
        \end{tikzpicture}%
    }
    \caption{Two limiting cases of tidal locking, at arbitrary scale. The solid purple ellipse is not locked to the host star (shown in yellow); the dashed purple ellipse is completely locked. These are the two cases that can currently be simulated using \texttt{greenlantern}.}
    \label{fig:tidally-locked}
\end{figure}

\bibliographystyle{aasjournal}
\bibliography{biblio}

\end{document}